\documentclass[lettersize,journal]{IEEEtran}
\usepackage{amsmath,amsfonts}
\usepackage{algorithmic}
\usepackage{algorithm}
\usepackage{array}
\usepackage[caption=false,font=normalsize,labelfont=sf,textfont=sf]{subfig}
\usepackage{textcomp}
\usepackage{stfloats}
\usepackage[hyphens]{url}
\usepackage{verbatim}
\usepackage{graphicx}
\usepackage{cite}
\usepackage{xcolor}
\usepackage{acronym}
\usepackage{multirow, multicol}
\usepackage{makecell}
\usepackage{caption}

\newacro{LLM}[LLM]{Large Language Model}
\newacro{ML}[ML]{machine learning}
\newacro{GPT}[GPT]{generative pre-trained transformers}
\newacro{MLM}[MLM]{masked language modeling}
\newacro{RL}[RL]{reinforcement learning}
\newacro{AGI}[AGI]{artificial general intelligence}
\newacro{RL}[RL]{reinforcement learning}
\newacro{SOTA}[SOTA]{state-of-the art}
\newacro{AR}[AR]{auto-regressive}
\newacro{NLP}[NLP]{natural language processing}
\newacro{VAE}[VAE]{variational autoencoder}
\newacro{NSP}[NSP]{next sentence prediction}
\newacro{SSL}[SSL]{self-supervised learning}
\newacro{AI}[AI]{artificial intelligence}

\begin{document}

\title{Wireless Multi-Agent Generative AI: From Connected Intelligence to Collective Intelligence}

\author{Hang Zou, Qiyang Zhao, Lina Bariah, Mehdi Bennis, and  M{\'e}rouane Debbah.
%\vspace{-15pt}
}
%\thanks{L. Bariah, Q. Zhao, Y. Tian, H. Zou, and F. Bader are with the Technology Innovation Institute (TII), 9639 Masdar city, Abu Dhabi, UAE, e-mail: firstname.lastname@tii.ae.}
%\thanks{M. Debbah is with the Department of Electrical and Computer Engineering, Khalifa University, Abu Dhabi, UAE, e-mail: merouane.debbah@ku.ac.ae.}

% \thanks{This paper was produced by by authors in Abu Dhabi, UAE and Oulu, Finland.}% <-this % stops a space
% \thanks{Manuscript received June 14, 2023; revised June 22, 2023.}
% \markboth{Journal of \LaTeX\ Class Files,~Vol.~01, No.~1, June~2023}%

% \IEEEpubid{0000---0000/00\$00.00~\copyright~2021 IEEE}
% % column for its text to clear the IEEEpubid mark.

\maketitle

\begin{abstract}
The convergence of generative large language models (LLMs), edge networks, and multi-agent systems represents a groundbreaking synergy that holds immense promise for future wireless generations, harnessing the power of collective intelligence and paving the way for self-governed networks where intelligent decision-making happens right at the edge. This article puts the stepping-stone for incorporating multi-agent generative artificial intelligence (AI) in wireless networks, and sets the scene for realizing on-device LLMs, where multi-agent LLMs are collaboratively planning and solving tasks to achieve a number of network goals. We further investigate the profound limitations of cloud-based LLMs, and explore multi-agent LLMs from a game theoretic perspective, where agents collaboratively solve tasks in competitive environments. Moreover, we establish the underpinnings for the architecture design of wireless multi-agent generative AI systems at the network level and the agent level, and we identify the wireless technologies that are envisioned to play a key role in enabling on-device LLM. To demonstrate the promising potentials of wireless multi-agent generative AI networks, we highlight the benefits that can be achieved when implementing wireless generative agents in intent-based networking, and we provide a case study to showcase how on-device LLMs can contribute to solving network intents in a collaborative fashion. We finally shed lights on potential challenges and sketch a research roadmap towards realizing the vision of wireless collective intelligence. 
\end{abstract}

%In magazine papers, index terms are not added usually ..
%\begin{IEEEkeywords}
%Multi-Agent System, Collective Intelligence, Large Language Models, Wireless Generative AI, 6G.
%\end{IEEEkeywords}

\section{Introduction}
% % Guidline of Mehdi
% \begin{itemize}
%     \color{blue}
%     \item LLMs are great and taking everyone by storn.....BARD, FALCON, GPT, babyGPTetc.
%     \item Taxonomy: System1 vs. System2
%     \item Most LLMs are system1; they hallucinate; can be constrained (using formal verification as done by your TII colleague)
%     \item What are their costs? Since our focus is on wireless: what are the tradeoffs?
%     \item LLMs also relate to protocol leaning \& semantic communication..
%     \item Use cases:? Human to machine; machine to machine; (see my slides)
% \end{itemize}
Generative \ac{AI} has evolved as a powerful branch of \ac{AI}, endowing machines with algorithms that enable them to create original content, such as images, text, and human-like conversations. In particular, \acp{LLM}, a particular model of generative \ac{AI} that are trained on a massive unlabeled textual dataset, have shown impressive capabilities in many applications, such as question answering, language understanding, reading comprehension, code generation, mathematical and common sense reasoning \cite{yenduri2023generative}. As a promising example of LLMs, the recent release of ChatGPT, based on the \ac{GPT} with 175B parameters (GPT-3) and 1 trillion parameters (GPT-4), has showcased remarkable capabilities of large-scale LLMs in content generation tasks. ChatGPT plugins allows \acp{LLM} to access up-to-date information, execute tasks, while interacting with the real-world.
% Meanwhile, many \ac{GPT} based \ac{LLM} emerge. 
More recently, the open-source Falcon \cite{falcon} has outperformed \ac{GPT}-3 with only 40B parameters, 75\% training cost, and 20\% inference compute budget; Similar results were achieved by LLaMA with 65B parameters \cite{touvron2023llama}. On the other hand, PaLM-2 model has excelled at advanced reasoning tasks including code generation and maths \cite{anil2023palm}. Such outstanding results of various LLMs has paved the way for integrating \ac{LLM} in different domains, such as robotics, telecom, and healthcare \cite{yenduri2023generative}.  
% Furthermore, \ac{GPT}-4 shows capability to interact with the real world, enabling intelligent agents to acquire and apply knowledge, solve problems, adapt to changing situations and achieve goals beyond their individual capabilities \cite{bubeck2023sparks}. 
The maturity of LLMs constitutes a stepping stone towards realizing the vision of \ac{AGI}, which is a broader concept than \ac{AI}, that encompasses highly autonomous systems of machines that possess general intelligence and cognition capabilities that are comparable to humans. Within this context, \acp{LLM} will play an essential role in \ac{AGI}-based systems through their abilities to perform complex tasks on multi-modal data across many domains, with only a few examples \cite{bubeck2023sparks}.

The remarkable capabilities of LLMs are owed to the Transformer architecture, rooted in the self-attention mechanism \cite{vaswani2017attention}. Leveraging multi-head attention, Transformers capture long-range dependencies in parallel by attending a word to all previous words, or to all other words in the text.  
%In most recent \ac{LLM}s, training \ac{GPT} with self-supervised learning on \ac{AR} next word prediction is used. 
With pre-training on huge unlabeled corpus, generative models can learn universal knowledge representation, and can be further tuned to narrow their scopes into a particular domain. Tuning generative LLMs on specific tasks or domains can be done via 1) fine-tuning by adapting pre-trained weights to task-dependent loss or domain-specific dataset; 2) prompt-tuning by applying task description or instruction with a few shot examples to help \acp{LLM} to understand the task; 3) instruct-tuning by \ac{RL} to ground \ac{LLM} into real-world context. 
% In many other \ac{LLM} (i.e. BERT), fine-tuning is widely used for downstream tasks on new domains. \ac{GPT} is proved to perform effectively with few-shot prompting, which significantly reduce computation cost. 

% on-device LLM

% Grounding, interactive, planning and reasoning
Accordingly, LLMs running on wireless devices should understand domain specific knowledge and interact effectively with the environment to perform specific tasks. Grounding LLMs in a real-world context with modular and causal knowledge is crucial to produce scalable and lightweight on-device LLMs. Such knowledge should be transferred in wireless networks and encoded into on-device LLMs, to perform logical decisions \cite{chaccour2022data}. This includes planning, that breaks down high-level goals into low-level tasks and reasoning, that deconstructs complex problems into priors and beliefs. Grounding, planning, and reasoning allow on-device \acp{LLM} to act as autonomous agents. 
% Grounding \ac{LLM} in a real-world context and interact with the environment is important for on-device LLM, which should understand domain specific knowledge of a wireless device and effectively perform the related tasks. 
% Grounding \ac{LLM} in a real-world context with knowledge, while enabling planning and reasoning is essential for on-device \ac{LLM}. The diverse environment where each wireless agent perceive and the tasks it performs require \ac{LLM} to understand, optimized, and modularized \cite{shen2023moduleformer} for the task and domain specific knowledge, in order to produce efficient, extendable, and specialized \ac{LLM} for lightweight deployment on devices. 
% Knowledge should be properly formulated in the concept space as abstraction of information. A wireless agent with \ac{LLM} should be able to leverage their accumulated knowledge and organize knowledge base to perform versatile and logical decisions \cite{chaccour2022data}. 
% multi-agent intelligence
Wireless networks represent a promising field for the deployment of generative multi-agent system, where multiple on-device \acp{LLM} plan and solve tasks in a collaborative manner. Specifically, multi-agent planning is essential to decompose a complex task that requires multi-modality sensors and multi-task executors across different wireless devices. Moreover, since on-device LLM encodes specific knowledge due to resource constraints, collective intelligence from multiple \acp{LLM} is essential to effectively complete complex tasks. To achieve this, generative multi-agent games can be applied with \acp{LLM} to solve advanced multi-task reasoning problems \cite{akata2023playing}. Furthermore, \acp{LLM} can be used with multi-agent \ac{RL} to learn optimal collaborative goal-oriented behaviours. %A knowledge-driven 6G network is essential for efficient multi-agent LLMs communication.  

% In this paper, we introduce a wireless multi-agent generative AI network, leveraging collective intelligence from massive connected devices to solve complex problems. Developing a computing fabric for 6G that moves knowledge within the network is essential for multi-agent LLM communication. 
% In doing so, 6G with collective intelligence can enable various applications in telecom, robotics, automobiles, manufacturing, healthcare, and so on. 

% discuss the architecture, theories, technologies and applications of a multi-agent generative AI system. Generative AI will be widely used on wireless devices. 
% Multiple generative agents can collaborate to solve a task or achieve a goal in various applications. In doing so, communication system should transfer effective information that leverage collective intelligence from multiple generative agents to accomplish the planned tasks. 
% 6G network will be a large scale distributed autonomous system powered by generative AI for various applications in telecom, robotics, vehicles, manufactory, and so on. 
\subsection{Related work}
The release of ChatGPT has led to a fast growing research in autonomous and generative agents. Several recently emerged frameworks expand on Chat\ac{GPT} from autonomous task planning and reasoning, to multi-agent generative systems. As an example, BabyAGI \cite{babyagi} is a task-driven autonomous agent built on \acp{LLM}, that can generate, execute and prioritize tasks in real-time. 
% , to perform a wide range of tasks across different domains. 
% It can complete tasks, generate new tasks from completed results, and prioritize tasks in real-time. 
% It stores completed tasks in a memory stream, which is used by \ac{LLM} to derive context for creation and execution of new tasks. 
Similarly, Auto-GPT \cite{autogpt} is an AI agent that attempts to achieve a goal specified in natural language.
% by breaking it down into sub-tasks and using a set of tools to execute in an automatic loop. 
It chains together \ac{LLM} "thoughts" in an infinite loop of reasoning, and planning the next action. 
% It adds on BabyAGI by giving \ac{LLM} the capability of choosing tools to execute the actions. 
AgentGPT is a user interface powered by the idea of BabyAGI and AutoGPT.
% \textcolor{red}{The idea of BabyAGI and AutoGPT are connected to the web as AgentGPT.} 
Another line of work  represented by HuggingGPT in \cite{shen2023hugginggpt} developed a collaborative system utilizing multiple AI models to complete a task.  In HuggingGPT, \acp{LLM} use language as an interface to connect numerous AI models in Hugging Face for solving complicated tasks. 
% The \ac{LLM} first plans a list of tasks based on user's request in text or image, assigns expert AI models to each task, then collects the results from experts and generate responds to the user. 
In this framework, \acp{LLM} act as a controller to manage and organize cooperation of expert AI models, which pushes the boundary of \acp{LLM} to applications requiring domain specific knowledge. % expert AI models, such as AI in radio communications. 

Furthermore, several frameworks on multi-agent LLM have been developed. A role-playing communicative agent framework called CAMEL has been developed in \cite{li2023camel}, in which an agent receives an idea to implement from a human, then creates an AI assistant and an AI user agent. 
% The roles of these two agents are assigned according to the specific idea, i.e. programmer and trader. 
The AI user assigns tasks for AI assistant to execute, and plans new tasks according to the results. The two agents collaboratively communicate by chatting with each other to solve the specified task. Moreover, a generative agent framework in a ``west world" simulation has been developed in \cite{park2023generative}. It is a sandbox of multiple AI agents that can interact with each other and simulate human behavior. The agent can observe the environment, plan a sequence of actions to execute, create high-level reflections of observation in a memory stream and formulate long-term plans. 
% store in a memory stream, plan a sequence of actions to execute. 
% The agent also retrieve relevant memory to form long-term plans, and to create higher-level reflections for future use. 
The agents communicate in full natural language using \acp{LLM}, to collaboratively complete a planned task. 

The \ac{SOTA} of generative agents explores multiple \acp{LLM} collaboratively planing and solving tasks. However, most works are executed in the cloud or in simulation environments, where the cost of communication, computing, and storage is ignored. In wireless generative agent networks, \ac{LLM} models and their inter-agent communication requires a new system design and optimization, which is the goal of this paper. 

\subsection{Contributions}

Motivated by the above discussion, in this article we establish the initial foundation for integrating the generative agents paradigm in wireless networks. In particular, we lay the first building block toward realizing collective intelligence in 6G, where we articulate the collaborative role of multiple on-device LLMs in solving Telecom tasks, and we put a future-oriented perspective on the architecture of multi-agent LLM-enabled wireless network. We further provide insights into the technologies that have manifested themselves as enablers for multi-agent LLMs in wireless networks, focusing on: 1) multi-agent LLM planning and reasoning to break down high-level goals into low-level tasks; 2) multi-agent LLM games and \ac{RL} to learn the optimal collaborative behaviours from competing actors to achieve a goal; 3) semantic communication that transfers knowledge in the network for system 2-type\footnote{System 2-type model refers to human-like cognition, based on reasoning and logical deduction, for solving complex problems and making calculated decisions.} on-device LLMs. With the aim to demonstrate its promising potential, we shed lights on the anticipated applications of generative agents in wireless networks, and we present a case study for showcasing a scenario where reasoning in multi-agent LLM is required to achieve a network-level energy saving goal and guaranteeing users' transmission rates. Finally, we explore future research directions for the successful realization of collective intelligence through on-device LLM.

\section{From Cloud LLM to Massive On-Device LLMs}
% % Guidline of Mehdi
% \begin{itemize}
%     \color{blue}
%     \item Need a nice table for comparing various LLMs (see earlier slide, GPT-3 paper Table D.1)
%     \item Edge AI \# of parameters, \# tokens; Energy bill? costs
%     \item Training and inference latency.; Architectures (are they all AR-TRANSFORMERs-based)?
% \end{itemize}
\subsection{Recent advances in cloud-based \ac{LLM}s}
 %{HangXXX: seems irrelevant to compare different \acp{LLM} here. Could be better to discuss the role of could LLM? A short glance of newest \acp{LLM} would be sufficient.}
% In this section, we will discuss several recently released \ac{LLM}s, focusing on their model size, training and inference cost, latency, and energy consumption. 
From an architectural perspective, the evolution of \acp{LLM} can be classified as encoder-only (BERT-like), decoder-only (GPT-like), and encoder-decoder (T5). The decoder-only \ac{GPT} merely utilizes the \ac{AR} transformer with multiple stacked transformer decoder layers, equipped with masked self-attention. This mechanism enables \ac{GPT}-like \acp{LLM} to attend all previous tokens to predict the subsequent one. GPT-4 (with plugs-in and web browsing) was reported to successfully solve the top 10\% of various academic and professional exams with improved reasoning capability.  %To facilitate the study of LLMs, OPT is published with parameter size ranging from tiny to huge. 175B-version OPT could maintain a comparable performance as GPT-3 while the carbon footprint requirement is merely 1/7 of the latter.  Recent released LLMs comes with even larger parameter size. e.g., GaLM, PaLM, LaMDA, Chinchilla, Jurassic-1, Gopher and MT-NLG. 
On the other hand, a number of lightweight \ac{LLM}s have been released, such as LLaMA and Falcon. These models have fewer parameters (Falcon-1B) and a smaller number of pre-trained tokens (1T), yielding the same performance as larger models \cite{falcon}. 

The encoder-only BERT is pre-trained on \ac{MLM} or \ac{NSP} tasks. Specifically, multiple tokens from a sequence are masked to force the model to acquire bidirectional contextual information during the pre-training process. The non-\ac{AR} nature of \ac{MLM} allows faster and paralleled computing by dynamically unfolding the masked tokens of all positions. 
% Diverse models enhanced, derived or extended from BERT have been proposed since its release. %To start with, Roberta largely improves the performance of BERT by optimizing some hyperparameters and training techniques. ALBERT reduce its number of parameters to 1/18 of BERT with almost equal performance. Similarly, sample-efficient ELECTRA is proposed by training a discriminative model which predicts whether the corrupted input is replaced. 
Enhanced versions of BERT can be achieved by optimizing the hyper-parameters and the pre-training tasks, i.e., RoBERTa and ALBERT, respectively.  
% BERT be easily extended to be language-specific (ERNIE for Chinese), multi-modal (ViLBERT and VideoBERT and be compressed in model size (DistilBERT, TinyBERT and Q-BERT). 
On the other hand, the encoder-decoder architecture, such as T5, converts diverse tasks that involve generating sequences or text into a text-to-text framework for pre-training. However, this architecture is computationally expensive to train, and has limited contextual understanding and reasoning. 
% compared to encoder-only or decoder-only ones. 

%The GPT framework has been extended to multi-modality. 
%Multi-modality \ac{LLM} has emerged after the success of GPT. 
Following the success of text-based \acp{LLM}, multi-modal \acp{LLM} have emerged to proliferate the potential use-cases of \acp{LLM}. Among others, DALL-E is a zero-shot text to image generative model \cite{ramesh2021zeroshot}, with 2 pre-training stages. During the first stage, a discrete \ac{VAE} is trained to compress image into grid of tokens. While in the second stage, the image tokens are concatenated with text tokens, in order to train an \ac{AR} transformer to model the joint distribution over the text and image tokens. With an enhanced approach, DALL-E2
% Furthermore, DALL-E2 \cite{ramesh2022hierarchical} shows better image generation by introducing CLIP latents. 
utilizes a CLIP similarity matrix to train a text encoder and an image encoder. The text embedding is then passed through a diffusion or \ac{AR} models prior to produce an image embedding and a diffusion decoder to generate images \cite{ramesh2022hierarchical}. A comparison of common \ac{LLM}s are given in Table \ref{tab:comparison}. 
% For energy consumption of training and inference compute, the largest one is considered for each model.

\begin{table*}[t]
	\centering
	\caption{Comparison of typical \ac{LLM}s}
	\label{tab:comparison} 
	\begin{tabular}{|c|c|c|c|c|c|c|c|c|}
		\hline
        \textbf{Model} & \textbf{Parameters} & \textbf{\makecell{Train data}} & \textbf{\makecell{Train energy}} & \textbf{\makecell{Inference FLOPs}} & \textbf{Architecture} & \textbf{Modality} & \textbf{Pretrain Tasks} \\ \hline
        GPT-3 & 175B & 300B tokens & 1287MWh & $7.4 \times 10^{14}$& Decoder & Text & MLM (AR) \\ \hline
        GPT-4 & 170T & 10T tokens & 7500MWh & ---  & Decoder & \makecell{Text, Image} & MLM (AR)  \\ \hline
        Falcon & 1B - 40B & 1T tokens & $\approx$960MWh & $\approx1.48\times 10^{14}$  &  Decoder & Text & MLM (AR)  \\ \hline
        LLaMA & 7B$-$65B & 1.4T tokens  & 449MWh  & --- & Decoder & Text & MLM (AR)    \\ \hline 
        PaLM-2 & 340B & 3.6T tokens  & --- & --- & Decoder & Text &  MLM (AR)   \\ \hline 
        T5 & 220M$-$11B  & 1T tokens & 86MWh  & --- &  Encoder-Decoder  & Text  &  \makecell{Text-to-Text \\ generation}  \\ \hline         
        BERT & 110M, 335M & 250B tokens & 1.536KWh & $3.4 \times 10^{11}$& Encoder & Text & \makecell{MLM, NSP}  \\ \hline
        RoBERTa & 110M, 335M & 2T tokens & $\approx$12.3KWh   & $3.4 \times 10^{12}$& Encoder & Text & \makecell{MLM, NSP} \\ \hline 
        DALL-E & 12B   & 2.5B pairs &  --- & --- & \makecell{dVAE, Decoder} & \makecell{Text, Image}  &  MLM (AR) \\ \hline       
        DALL-E2 & 3.5B & 6.5B pairs & ---  & --- & \makecell{Encoder CLIP \\ AR / Diffusion Prior}  & \makecell{Text, Image} & \makecell{Text-to-Image \\ diffusive generation } \\ \hline        
        % Whisper & 39M$-$1550M  & \makecell{680k hours} & --- & --- & Encoder-Decoder &  \makecell{Text, Audio}  &  Seq-to-Seq  \\ \hline 
	\end{tabular}
\end{table*}

\subsection{On-device \ac{LLM}s Challenges}
Despite their promising capabilities, building intelligent wireless networks that are empowered by generative agents is extremely challenging using current \acp{LLM}. This is due to their huge parameter sizes, which limits their deployment at the edge. In particular, it was demonstrated that, in few-shot scenarios, models' memory requirements dramatically increase as the number of parameters increase. 
For example, it takes roughly 86 GB of GPU memory for Falcon-40B to infer. 
%While the existence of emerging abilities of \acp{LLM}
% , i.e., abilities not present in smaller-scale models that are present in larger-scale models, 
%is still under debate, their capacities especially in few-shot scenario skyrocket as the scale of parameter size increases. 
% For example, in-context learning, i.e., learning from a few examples in the context is generally observed for LLMs with billions of parameters e.g., instructGPT, GPT-3 and PaLM. 
In addition to that, \acp{LLM} have long inference time when solving sophisticated tasks (planning and reasoning), which significantly increases the communication latency. 
% long inference time of \acp{LLM} facing sophisticated tasks such as planing and reasoning would significantly increase the latency of communication. 
Finally, updating or knowledge synchronization from central \acp{LLM} to on-device \acp{LLM} could be energy-consuming through conventional techniques, such as federated learning. 
% Therefore, \acp{LLM} which can operate normally in the ``on-device" scale of parameter size are crucial for connected intelligence with enormous amount of devices.

With the aim to realize lightweight LLMs, neural network compression techniques such as weight sharing, pruning, knowledge distillation, low-rank decomposition, and quantization can be applied. Within the context of compression, it should be highlighted that, since \acp{LLM} are few-shot learners rather than task-specific models, compression techniques should preserve LLM's transferring abilities. In a different context, QLoRA is a high-precision quantization technique, that can be used to quantize an LLM to 4-bit, then adds a small set of learnable low-rank adapters, which are tuned through back-propagating gradients using the quantized weights. This significantly reduces the GPU memory and latency in fine-tuning and inference. Although compression and quantization can reduce the model size and computation requirements, due to the non-modular nature of LLMs' knowledge, such approach can yield degraded performance in specific tasks. 
% State-of-the-art chatbot performance could be achieved by fine-tuning various \acp{LLM} with significantly less GPU memory and training time compared to the regular fine-tuning techniques. 
% Shall we focus more on the network side, transmitting data instead of knoweldge?

\section{Multi-Agent Collective Intelligence via Generative LLM}

\subsection{Multi-Agent LLM Network Architecture}
\begin{figure*}[h!]
\centering
\includegraphics[width=1\linewidth]{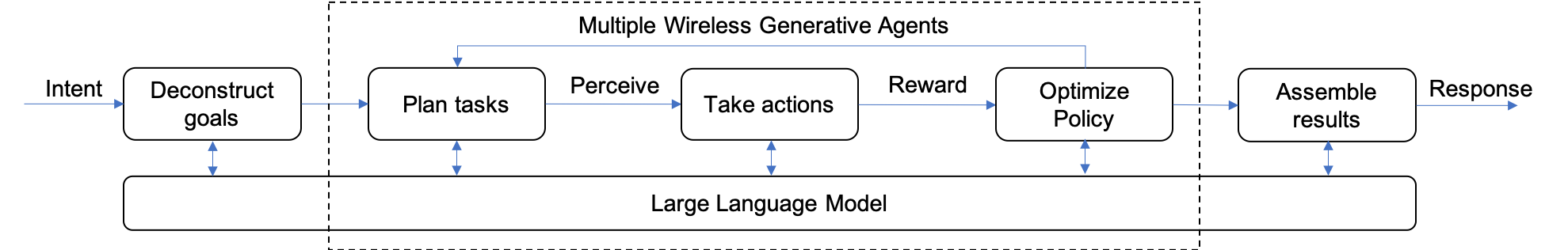}
\caption{Close-loop task planning, execution, optimization in wireless generative agents.}
\label{fig:example}
\end{figure*}

Deploying generative AI in wireless networks requires the realization of collective intelligence from multiple on-device LLMs, that can interact with the environment, plan actionable tasks from a goal, and solve the tasks collaboratively by exchanging knowledge. Our aim in this section is to set the groundwork for integrating generative agents into future wireless networks, in which a wireless generative agent leverages LLMs to perform a close-loop task planning, execution, and optimization. As illustrated in Fig. \ref{fig:example}, this process can be achieved through multiple steps. First, the agent deconstructs higher-level intents or goals, and plan sequences of lower-level tasks over time. Afterwards, it perceives the environment, performs decisions and takes actions accordingly, and then optimizes the decision policy from the rewards. According to the results, the agent creates new tasks until the intent's goal is achieved, and then generates a final response. 
% \begin{itemize}
%     \item Deconstruct higher-level intents or goals, to plan sequences of lower-level tasks over time;
%     \item Perceive environment and generate actions to execute, and optimize decision policy from rewards;
%     \item Create new tasks until the intent's goal is achieved, and generate response in a requested format;
%     % \item It connects LLM with actors to execute instructions generated by LLM, retrieve results or feedbacks from environment, and use LLM to plan next actions; 
%     % \item It uses LLM to create abstracted thoughts of observations and actions taken by other agents, retrieve the relevant and important information for the planned tasks;
% \end{itemize}

\begin{figure*}[h!]
\centering
\includegraphics[width=1\linewidth]{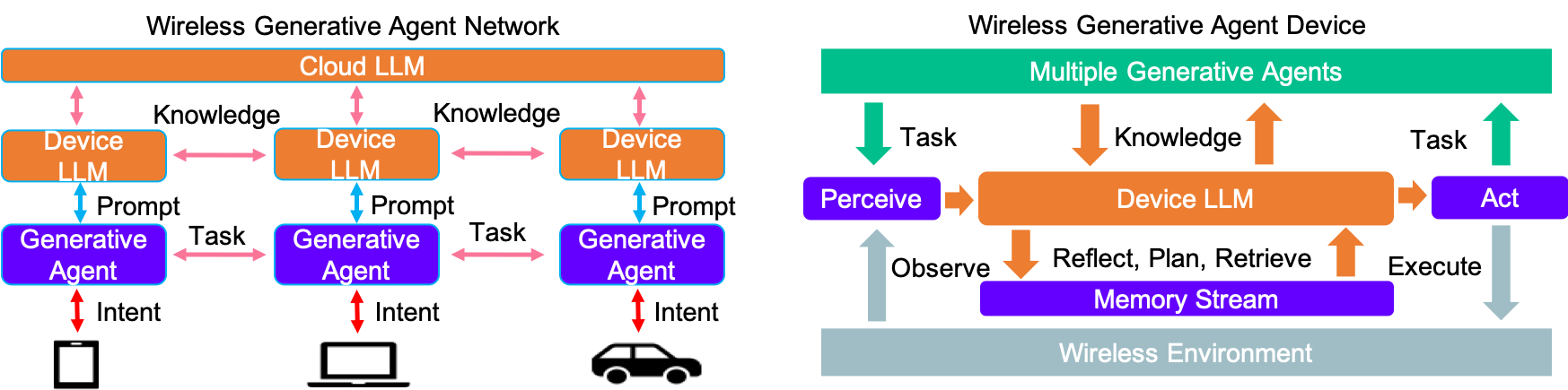}
\caption{Wireless generative agent network and device architecture.}
\label{fig:architecture}
\end{figure*}

In order to realize collective intelligence, the earlier mentioned procedure can be performed through multiple wireless generative agents, in which the agents exchange knowledge through the network, to collaboratively plan tasks, take actions and optimize policies, based on the architecture shown in Fig. \ref{fig:architecture}. Note that knowledge is an abstracted representation of the data, which can be encoded into LLMs for performing specific tasks. From the network perspective, intents from humans or machines are provided to generative agents through different wireless terminals. These terminals create prompts to an on-device LLM to complete each step, as shown in Fig. \ref{fig:example}. The tasks are planned collaboratively among multiple generative agents, to best leverage the knowledge of different LLMs, and the capabilities of different devices. The on-device LLM can obtain domain specific knowledge from the cloud-based LLM, or from other on-device LLMs when the planned task is received from other agents. From the device perspective, a wireless generative agent has a perceiver (sensor) to observe the environment, and an actor (controller) to execute the decisions. The on-device LLM extracts semantic information from the observed raw data in multi-modalities (text, image, sound), and stores it in a memory stream for planning new tasks in the future. Accordingly, to perform a particular task, the relevant semantic information is retrieved for taking an action. After completing its planned actions from the received higher-level tasks, the agent can further create lower-level tasks and send it to other agents to complete the goal. 
% deconstruct high-level intents from human or machines, plan sequences of tasks over time, perceive environment and represent with abstracted thoughts, take actions and optimize decision policy from environmental rewards, create new tasks until the intent's goal is achieved, and finally generate response in a requested format. 
% It leverages LLM to plan sequences of actions over time, reflect abstracted thoughts of environment observations, prioritize and instruct executors to take actions, analyze results or feedbacks and create new necessary tasks until the initial goal is achieved, and generate responses in the formats as desired from the prompts, such as text, image, or codes. 
% A wireless generative agent includes close-loop task planning, reasoning and knowledge management (Fig. \ref{fig:example}). 
% \begin{itemize}
%     \item It automates the use of LLM in solving complicated tasks by extracting LLM's responses and continuously creating new prompts to achieve the task goal; 
%     \item It connects LLM with actors to execute instructions generated by LLM, retrieve results or feedbacks from environment, and use LLM to plan next actions; 
%     \item It uses LLM to create abstracted thoughts of observations and actions taken by other agents, retrieve the relevant and important information for the planned tasks;
% \end{itemize}

% Wireless generative agents perceive their environment and save in a memory stream. Based on their perceptions, the architecture retrieves relevant memories, then uses those retrieved actions to determine an action. These retrieved memories are also used to form longer-term plans, and to create higher-level reflections, which are both stored the memory stream for future actions (Fig. \ref{fig:architecture}).

% The agents communicate with reflected perceptions and planned actions, in a form of abstracted semantic information generated by LLM. This can be natural language or embedding on latent space, which is understandable by the LLM in received agents. Furthermore, the LLMs communicate with knowledge to learn from each other for better understanding different scenarios and solving complicated tasks. The agent can either learn from cloud LLM with world knowledge, or other agents in network having more expertise in the specific (sub)task. 

\subsection{Multi-Agent LLM Technologies}

\subsubsection{Planning and Reasoning}
% Guidline of Mehdi
%\begin{itemize}
    %\color{brown}
    %\item Dynamic chain, trees, graph of thoughts, actions, tasks. Planning actions is a base of Auto-GPT, and to realize intent based X.
%\end{itemize}
A high-level goal for wireless agents usually contains complex tasks or problems, and therefore, planning is an essential step for the generative agents to create actionable tasks over time to achieve a specific goal. The currently available \acp{LLM} have limited capabilities in solving problems with complex reasoning and multi-step planning. 
% For example, PaLM (540B) with standard prompting could merely achieve a success rate of $18\%$ on math problems at the grade school math level of GSM8K dataset. 
This can be a limiting factor in the deployment of wireless generative agents in networks with ultra-high reliability requirement. 
% Poor reasoning results of accuracy-sensitive tasks, e.g., resource allocation, could be challenging for operating a reliable generative agent network. 
Therefore, system 2-type \acp{LLM} can endow wireless generative agents with reasoning capabilities. Recently, many works have shown the reasoning abilities of \acp{LLM} through prompting engineering. For instance, chain-of-thought prompting, which prompts \acp{LLM} with a sequence of short sentences serving as intermediate reasoning steps to solve problems, have demonstrated an outstanding performance on a wide range of reasoning tasks. Thereafter, self-consistency strategy is proposed to replace the greedy decoding of chain-of-thought prompting. Furthermore, tree of thoughts is another method which maintains a tree for each immediate step of thought so that  \acp{LLM} can deliberately evaluate multiple paths and adjust decisions as needed for optimal outcomes. As an alternative to prompts optimization which can be regarded as an outcome-supervision, one possible solution is to supervise along the reasoning process. This requires LLMs to properly model and encode a domain specific knowledge for solving different tasks. Furthermore, multi-agent planning can reduce the cost of inference or fine-tuning, by distributing the task between multiple LLM devices with domain specific knowledge. 

\subsubsection{Multi-agent LLM Games and RL}
% Guidline of Mehdi
%\begin{itemize}
    %\color{brown}
    %\item Multiple agents cooperatively complete the planned actions under competition environments (resources, multi-intents, multi-plans)
%\end{itemize}
Current \acp{LLM} are trained to generate contents which are helpful, reliable, and harmless to the users. These attributes can be achieved only if one-to-one iteration between a human and an agent is considered. However, in the presence of multiple generative agents, it is yet debatable whether these features can be realized. For instance, egocentric agents, which are extensively serving their corresponding users, can hinder the normal operation of other agents or be unresponsive to some collaborative tasks. On the contrary, selfless agents which are instructed to be helpful can be vulnerable to malicious attacks. Hence, it is essential to strike a balance between selfishness and selflessness when designing generative agents. Within this context, game theory constitutes a suitable tool to model multi-agent LLMs and analyze their behaviour. Initial attempts to understand the interactions between multiple \acp{LLM} (i.e.  \textit{\ac{LLM} games}) are presented in \cite{akata2023playing}.

Multi-agent \ac{RL} (MARL) is another approach to observe the optimal collaborative behavior in a multi-agent system, which can also be applied to LLMs. In particular, it is essential for the wireless agents to interact with the environment, where \ac{RL} tunes the actions generated by \acp{LLM} with environment rewards, in order to ground  \acp{LLM} into a real-world context. MARL-enabled LLMs can further model the interaction between the wireless agents, using LLM's knowledge, to learn the optimal collaboration policy and communication protocols among distributed LLM agents. In doing so, communication costs can be reduced as agents converge towards optimal decision policies. % in different states, which can perform more local decisions.

\subsubsection{Semantic Information and Communication}
% Guidline of Mehdi
%\begin{itemize}
    %\color{brown}
    %\item Semantics of information accumulated as knowledge in long term, to store in the device memory and exchange between devices for LLM to take the planned actions
%\end{itemize}
The interaction between the multiple wireless generative agents in collaborative systems will result in a huge amount of information to be generated, exchanged, and perceived. Shannon communication rooted in the level A of Weaver's 3-level communications solely aims at transmitting symbols accurately and efficiently, while ignoring the conveyed meaning of the transmitted messages. In contrast, in semantic communication, the transmitter sends only useful/relevant information to the receiver, in order to solve a given downstream task. This paradigm is possible if the goal/objective of the communication is known at both sides. Recalling that wireless generative agents aim to collaboratively solve a specific problem, semantic communication can transfer task specific knowledge for LLMs on targeted devices. Furthermore, agents can communicate based on their abstracted state information, to improve multi-agent cooperation with efficient protocols. 
% A naive semantic encoder for an agent could be a simple summarizing prompt or question-answering prompt to its supporting \acp{LLM}. Another example of semantic-level communication could be a briefing from one agent to inform related agents about its impacting actions or state information. Obviously this tiny  information is efficient compared to the situation where other agents have to estimate or sense this information.

Similarly, semantic information can enable LLMs to integrate multi-modal raw data in a common concept space. In doing so, the data exchanged and stored among wireless devices can be largely reduced. Moreover, LLMs can exploit abstracted "thoughts" from the perceived data by modelling it using semantic information on a topological space. Hence, LLMs can perform effectively in short-term actions with lower-order information, and in long-term plans with higher-order abstractions. This also enables LLMs to encode domain or task specific knowledge, and thereby, to reduce the model size and computing costs, making it suitable for resource constrained devices. 

%\subsubsection{Federated Learning}

% \subsubsection{Emerging Protocol Learning}
% \begin{itemize}
%     \color{brown}
%     \item Emerging L2 protocol to coordinate communication between agents to achieve intent’s goal.
% \end{itemize}

\section{Applications of Multi-Agent LLM Networks}
% \subsection{Multi-Agent LLM Applications}
\subsection{Intent-driven Network Automation}
% \subsubsection{Intent-driven Autonomous Networks}
With the vision to enable network automation in future wireless networks, it is envisioned that 6G will support intent-based networking, which aligns the network operation with particular intents and objectives. Although intent-driven networks are defined in 3GPP and IETF, their functionalities and their flexible deployment are still limited. On the one hand, intent translation and orchestration relies heavily on services models and policies, which limits its capabilities in handling new scenarios. On the other hand, intent driven networking has not 
been applied yet in general networks and user devices, including radio access networks. 
%Intent-driven networking is an important 6G use case, which has been developed specifically in 5G network slicing, where service providers define an expectation in natural language, such as “improve network quality”, then use AI to translate it into real-time network actions, where the network continuously monitors and matches the intent with minimal human intervention. 

% is limited to specific domains of network,
% such as operation in transport or core network. It has not yet been widely adapted to general network devices (gNBs, UEs) or user devices (mobile, machine). 
% Furthermore, communication networks and applications are still decoupled, making it difficult for the intent’s goal to be achieved by an optimized network. 

LLMs can potential extend the concept of intent driven networks to a general purpose self-designed communication network. The Open AI’s ChatGPT-4 has recently been equipped with a number of plugins, allowing an autonomous agent to execute the commands generated by GPT and produce outputs in natural languages. Inspired by this, intent driven networking can utilize LLMs to break down higher-level intents, (e.g., reducing the network energy consumption by 5\%) into a sequence of lower level tasks (e.g., tuning transmit power, channel measurement, etc.), and instruct related executors to take actions (i.e. gNB). The results are then prompted to the LLM, which will add subsequent tasks, if necessary, until the intent’s goals are achieved. In a multi-agent LLM network, each network device will leverage LLMs to analyze the perceived information, including the observations from the environment and actions taken by other agents. Once the agent plans and prioritizes a set of sub-tasks, it takes actions on local tasks, which can be accomplished by the local actor. After that, it utilizes LLMs to generate a protocol language including the remaining and newly created sub-tasks, and sends to other agents through the wireless network. Each agent conducts such loops and communicates with others until the added tasks are completed and the goal is achieved.

\subsection{Case Study: Wireless Energy Saving}

% \section{Wireless Generative Agents in Intent-based Networking: A Case Study}

% % Guidline of Mehdi
% \begin{itemize}
%     \color{blue}
%     \item Results for 1-2 use cases (ideally different ones; for edge AI LLM and more multi-agent robot-style or baby-GPT style ones). 
% \end{itemize}
{The vision of generative agent-enabled intent-driven networking requires LLMs to deconstruct a wireless network goal into sub-tasks or sub-problems and solve them via multi-agent cooperation. One important step in this process is the efficiency of multiple LLM instances in formulating and solving problems pertinent to wireless networks. In this section, we consider a scenario with multiple mobile users playing a game to reduce the total power consumption. Our aim is to evaluate the capability of on-device LLMs in solving a wireless communication problem, and show their potentials in enabling an end-to-end autonomous network.
}

% In this section, we present an example of multi-agent LLMs solving an energy saving goal in a wireless network. Inspired by the multi-agent LLM game in \cite{akata2023playing}, we experiment with GPT-4 assuming a scenario with multiple mobile users playing a game to reduce the total power consumption. The objective of this experiment is to evaluate the efficiency of multiple LLM instances in formulating and solving mathematical problems in a wireless network.

In our setup, we assume a network with $K = 4$ users transmitting in a shared spectrum. Each user has a channel gain of $g = \left(1.21, 2.01, 0.58, 0.13 \right)$, bandwidth $b = 15\text{kHz}$, and noise $n = 1\text{dB}$. The game starts with an initial transmit power $p = \left(2, 4, 5, 6 \right) \text{W}$. The base station has the goal of "reducing network energy consumption by $\Delta p = 0.85$ W" while 
maintaining individual transmission rates above $r = \left(3.50, 15.80, 4.40, 1.00 \right) \text{Kbps}$, respectively. We create independent LLM instances for each user. The above environment information is given to each LLM instance in the beginning of the game. Moreover, the users compete for power savings goal under mutual interference. {Note that when implemented in a real wireless network, these parameters are measured by on-device perceivers and provided to on-device LLMs.} We generate our results using the GPT-4 model.
\begin{figure*}[h!]
\centering
\includegraphics[width=1\linewidth]{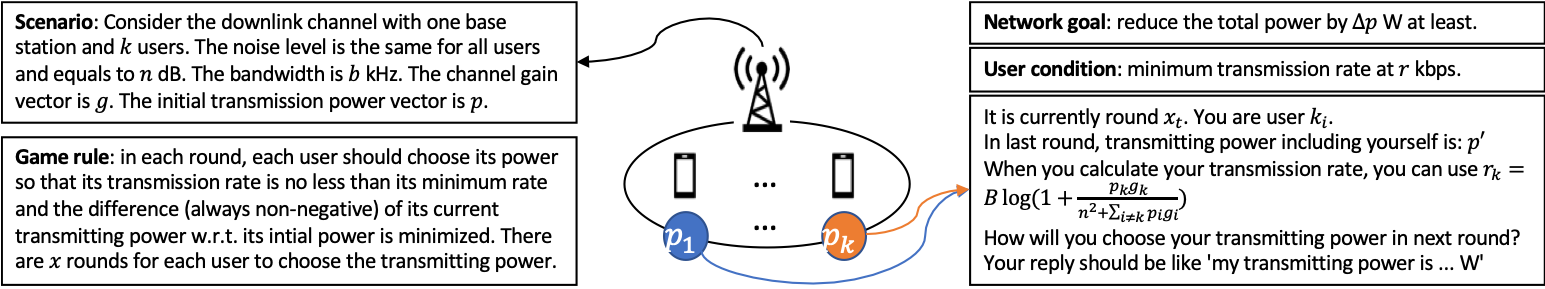}
\caption{Playing repeated games in an example of power allocation on mobile users to minimize network power consumption.}
\label{fig:game}
\end{figure*}

\begin{figure}[h!]
\centering
\includegraphics[width=1\linewidth]{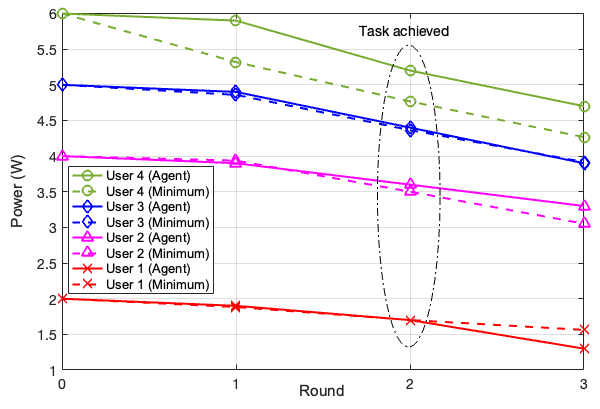}
\caption{Users' transmission power given by LLMs, and power to maintain a minimum transmission rate  v.s. round for reducing total power by $5\%$ at least. }
\label{fig:power_history}
\end{figure}

\begin{figure}[h!]
\centering
\includegraphics[width=1\linewidth]{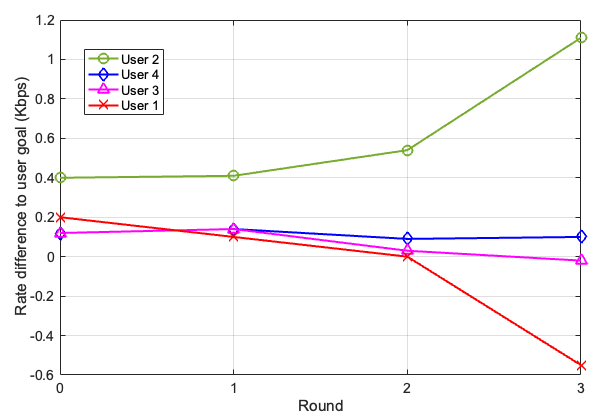}
\caption{The difference of transmission rate w.r.t. minimum rate  v.s. round for total power reduction by at least $5\%$. }
\label{fig:rate_diff_history}
\end{figure}

The prompts we created for LLMs to play this game are shown in Fig. \ref{fig:game}. In the beginning, a prompt consisting of general description, game rules, goal of the base station, and the individual goal of each users is passed to each LLM instance. In each round, the LLM instances are requested to choose a new power level $p_i$, given the actions of all users in the previous round, through a prompt.
% Our simulation setting is the following: channel gain $g = \left(1.21, 2.01, 0.58, 0.13 \right)$; noise level $n = 1\text{dB}$; initial transmission power $ p = \left(2, 4, 5, 6 \right) \text{W}$; minimum rates are $r = \left(3.50, 15.80, 4.40, 1.00 \right) 
% \text{Kbps}$;  bandwidth is $b = 15\text{kHz}$. 
The transmission powers chosen by each agent from GPT-4 in each round are illustrated in Fig. \ref{fig:power_history}. The obtained results are compared with the minimum theoretical power required for the transmission rate $r$. It can be seen that LLM-enabled agents managed to achieve the power saving target $\Delta p$ at round 2, with 3 users allocated the theoretical minimum power to keep their transmission rate at above $r$ (except user 1). This can be further noticed in Fig. \ref{fig:rate_diff_history} which shows the users' rate difference to $r$, where user 1's rate can be further reduced, and hence, the transmission energy can be reduced. Furthermore, users 1 and 4 violate the goal of minimum transmission rate in round 3. The simulation results in Fig. \ref{fig:rate_diff_history} also show that current GPT-4 experiences some difficulties in maintaining multiple goals when the number of agents increases. 

This use-case demonstrates that an LLM (GPT-4) can perform mathematical reasoning in a wireless communication-based problem, in order to achieve a global power saving and individual transmission rates targets. 
%The strategy of all agents in each round is cooperative even if agents are asked to maintain their rates as much as possible. The reason for this behavior could be the requirement during instruct-tuning, i.e., LLMs are trained to assist human requests as much as possible. 
The takeovers of this study are: 1) LLMs have the potential to analyze a radio environment, formulate and solve a radio network problem to achieve a network operation goal; 2) Multi-agent LLMs can achieve a cooperative goal while keeping their own KPIs; 3) Training a wireless domain-specific LLM is essential for LLMs to understand effectively the wireless network goals (without mathematical prompts), to achieve a fully autonomous intent-driven network.

% The difference between transmission rate w.r.t. minimum transmission rate in each round is illustrated in Fig. \ref{fig:rate_diff_history}. 

% \textcolor{brown}{
% We are experimenting a use case where multiple \ac{LLM}s collectively solve an energy saving problem in wireless network. 
% \begin{itemize}
%     \item In brief, consider network users have their own LLM instance, we first prompt with environment information (bandwidth, noise, initial power, …), then we give a goal (reduce total energy by 5W) and a rule (minimum transmission rate 2 kbps). We ask each LLM to play an action (choose optimal power), and give a reward (power / capacity of other users). 
%     \item This very simple use case could demonstrate the capability of LLM: 1) play multi-agent games and reinforcement learning; 2) planning: formulate a mathematical problem from a goal in wireless; 3) math reasoning: solve wireless tasks from math instead of training a (radio) data based model (the way in our other application). 
% \end{itemize}}
% {Mehdi: try to collect as many KPIs as possible (to show various tradeoffs, latency, energy, etc) but also the ML ones (no. of token, inference time, and so forth). This big picture is needed.}

% \begin{figure*}[h!]
% \centering
% \includegraphics[width=1.0\linewidth]{Figure/beam_power_task.png}
% \caption{A example of multiple LLM agents implemented on wireless devices to collaboratively complete an intent of beam and power optimization. The tasks are generated by Auto-GPT with refinements. LLM instructs different actors on RIC, BS and UE to execute the planned task, and generate response of the achieved KPI.}
% \label{fig:example_intent}
% \end{figure*}

\section{Challenges and Opportunities}

\subsection{TelecomLLM with domain knowledge}
Building a TelecomLLM with domain-specific knowledge is a foundation to enable wireless generative agents. Current GPT-4 can only provide general responses to a wireless network goal described in natural language. To operate an autonomous network, the LLM should give instructions to specific network functions or entities. This requires LLMs to be trained on Telecom domain corpus, such as standard and product specifications. Moreover, grounding and calibrating on-device LLM to real-time wireless environment is important for wireless agents to make optimal decisions on specific wireless tasks. Furthermore, tuning TelecomLLM with emerging wireless knowledge is crucial to keep the model updated with new wireless standards and features, in order to support efficient network upgrade. 

\subsection{Wireless LLM Hallucination}
Model hallucination phenomenon refers to the event when the generative model produce content that is nonsensical or untruthful in relation to certain sources. In wireless networks, this can be a result of continuous channel variations and network dynamicity. In particular, model hallucination can be experienced when the generative agent is not capable of capturing the variations of a rapidly evolving network, leading to inaccurate representation of the wireless environment. Furthermore, generative agents may hallucinate due to the impact of inter-agent interference and the lack of relevant on-board data. Hence, efficient hallucination avoidance schemes are needed to ensure trustworthy agent behaviours in different network functionalities. 
\color{black}

\subsection{Self-replicating Wireless Generative Agents}
The concept of self-replication within the context of generative agents indicates their capability of creating copies of themselves, including their knowledge and experiences, to communicate effectively with other agents. While still being a speculative and a hypothetical concept, it represents a promising skill that enables such agents to cope with the growing nature of wireless networks, and achieve autonomously scalable wireless networks. Also, replicated agents can adapt their models to their own wireless environments and tasks. Knowledge driven wireless generative agents and inter-agent communications are key enabling technologies for self-replicating agents. For example, modularized LLMs have the potential to provide flexible knowledge inheritance for different generative agents, with reduced training and communication overhead. This calls for a thorough study in explainable LLM which can encode knowledge on-demand. 
It is worthy to emphasize that, although self-replicating generative agents are aimed for improved wireless networks performance, reliable wireless communication is essential for ensuring accurate knowledge transfer, and therefore, to prevent replication errors and agents mutations.
\subsection{Resource management for on-device LLMs}
The inherent limitations on the available resources in wireless networks, particularly IoT systems, represent a performance bottleneck in the efficient realization of generative multi-agent systems in future wireless networks. While being a promising technology, due to the diverse capabilities of wireless nodes, on-device LLM requires a sophisticated balance between computing and energy resources, memory constraints, inference reliability, as well as user's experience and demands. Accordingly, it is essential to develop optimization and resource-management approaches that can be tailored to a certain use-case, through assessing the resource utilization performance of each agent, and its impact on the CPU, memory, and energy efficiency, while meeting particular requirements imposed by the communication scenario. 

\subsection{Multi-Agent Convergence}
Multi-agent convergence in generative systems can be particularly challenging in wireless networks, due to the dynamic nature of the latter, and therefore, the continuous need for agents' models adaptation to cope up with the evolving network conditions. Such a challenge is particularly pronounced when a massive number of agents are involved in achieving different network goals. Failing to converge might result in suboptimal solutions, which will then necessitate a high level of coordination among the multiple agents, contradicting the goal of achieving self-governance and automation among the agents. Accordingly, this opens up a new horizon for exploring collective approaches to allow the agents to achieve a consensus in a real-time manner, regardless of the network dynamicity condition. 
%when agents have different goals, and the number of agents increase. The agents may require more exploration rounds to achieve the target, or may not find an optimal solution. This can cause computing and communication cost. Furthermore, it is important for agents to identify a proper time to stop interaction, and stay on a stable solution. 
% For example, in our experiment, further rounds of game could on a contrary destroy the achieved goal. 
%Optimizing the cooperative planning of tasks among multiple agents by considering changes of each agent's behaviour and convergence over time can improve the system level convergence.

\subsection{Multi-modal efficient on-device LLMs}
% While LLMs are mainly focusing on textual and visual data, 
Generative agents employed in wireless networks are anticipated to handle multi-modal data, such as text, images, locations, and radio signals. {Meanwhile, multi-modal on-device LLM is constrained by communication and computing resources. This raises significant challenges in the model design, where current multi-modal LLMs encoding raw data are too large to be deployed on mobile device. Knowledge based LLM that encodes multi-modal data from a concept space can potentially reduce the model size and inference cost, by learning a minimal structure of data. Research efforts should be devoted to build effective concept space for multi-modal data.
% For example, a multi-modal learning framework creates 2D projection of 3D point-clouds, to perform feature fusion with images through ResNet. This significantly reduces computation compared to PointNet. However, it also cause large information loss and has limited applicability. Thus, research efforts should be devoted to build effective concept space for multi-modal LLMs. 
} 

% This raises several challenges pertaining to Telecom data availability for training and fine-tuning. In particular, on-device LLM are specifically vulnerable to the availability of relevant datasets, for models fine-tuning, with the aim to achieve a particular goal related to specific network requirements. Furthermore, LLMs deployed on edge devices are expected to be adaptive enough to accommodate the varying network requirements, while being limited in terms of available data (due to memory and resources constraints). In addition, generative agents are expected to perform efficient fusion for the different data modality, in order to allow the agents to enjoy improved contextual and situational awareness, and therefore, enable them to perform efficiently in different network scenarios. 

\section{Conclusion}
In this paper, we introduced the concept of multi-agent generative AI network, which exploits the collective intelligence from on-device LLMs. We identified the challenges and key enabling technologies of on-device LLMs, focusing on LLM planning and reasoning to solve complex tasks; multi-agent LLM games and \ac{RL}-based LLMs to achieve goals in competition scenarios; and semantic information and communication to connect LLMs with knowledge for effective short-term and long-term task planing. Moreover, we discussed potential applications of multi-agent LLMs in 6G, including intent-driven network automation. We further demonstrated a use case where multi-agent LLMs can play a game to achieve goals of network energy saving and user transmission rate. Finally, we discussed potential research opportunities of multi-agent LLM network such as system 2 ML with strong reasoning, human-agent interaction and collaboration. This paper initiates a new framework for future wireless network design to empower collective intelligence from large scale wireless devices, opening up new research opportunities of generative AI in wireless networks. 

\bibliographystyle{IEEEtran}
\bibliography{ref}

\begin{thebibliography}{10}
\providecommand{\url}[1]{#1}
\csname url@samestyle\endcsname
\providecommand{\newblock}{\relax}
\providecommand{\bibinfo}[2]{#2}
\providecommand{\BIBentrySTDinterwordspacing}{\spaceskip=0pt\relax}
\providecommand{\BIBentryALTinterwordstretchfactor}{4}
\providecommand{\BIBentryALTinterwordspacing}{\spaceskip=\fontdimen2\font plus
\BIBentryALTinterwordstretchfactor\fontdimen3\font minus
  \fontdimen4\font\relax}
\providecommand{\BIBforeignlanguage}[2]{{%
\expandafter\ifx\csname l@#1\endcsname\relax
\typeout{** WARNING: IEEEtran.bst: No hyphenation pattern has been}%
\typeout{** loaded for the language `#1'. Using the pattern for}%
\typeout{** the default language instead.}%
\else
\language=\csname l@#1\endcsname
\fi
#2}}
\providecommand{\BIBdecl}{\relax}
\BIBdecl

\bibitem{yenduri2023generative}
G.~Yenduri \emph{et~al.}, ``Generative pre-trained transformer: A comprehensive
  review on enabling technologies, potential applications, emerging challenges,
  and future directions,'' 2023.

\bibitem{falcon}
Falcon {LLM}. \url{https://falconllm.tii.ae}.

\bibitem{touvron2023llama}
H.~Touvron \emph{et~al.}, ``{LLaMA}: Open and efficient foundation language
  models,'' 2023.

\bibitem{anil2023palm}
R.~Anil \emph{et~al.}, ``{PaLM} 2 technical report,'' 2023.

\bibitem{bubeck2023sparks}
S.~Bubeck \emph{et~al.}, ``Sparks of artificial general intelligence: Early
  experiments with {GPT}-4,'' 2023.

\bibitem{vaswani2017attention}
A.~Vaswani \emph{et~al.}, ``Attention is all you need,'' 2017.

\bibitem{chaccour2022data}
C.~Chaccour \emph{et~al.}, ``Less data, more knowledge: Building next
  generation semantic communication networks,'' 2022.

\bibitem{akata2023playing}
E.~Akata \emph{et~al.}, ``Playing repeated games with large language models,''
  2023.

\bibitem{babyagi}
Y.~Nakajima, ``Task-driven autonomous agent utilizing {GPT}-4, {Pinecone}, and
  {LangChain} for diverse applications,'' 2023.

\bibitem{autogpt}
Auto-{GPT}: An autonomous {GPT}-4 experiment.
  \url{https://github.com/Significant-Gravitas/Auto-GPT}.

\bibitem{shen2023hugginggpt}
Y.~Shen \emph{et~al.}, ``Hugginggpt: Solving {AI} tasks with {ChatGPT} and its
  friends in {HuggingFace},'' 2023.

\bibitem{li2023camel}
G.~Li \emph{et~al.}, ``{CAMEL}: Communicative agents for "mind" exploration of
  large scale language model society,'' 2023.

\bibitem{park2023generative}
J.~S. Park \emph{et~al.}, ``Generative agents: Interactive simulacra of human
  behavior,'' 2023.

\bibitem{ramesh2021zeroshot}
A.~Ramesh \emph{et~al.}, ``Zero-shot text-to-image generation,'' 2021.

\bibitem{ramesh2022hierarchical}
A.~Ramesh, P.~Dhariwal \emph{et~al.}, ``Hierarchical text-conditional image
  generation with clip latents,'' 2022.

\end{thebibliography}
% \nocite{*}

% \end{IEEEbiography}

% \section*{Biographies}
% \textbf{John Dose} (john.dose@gatech.edu) is with the Department
% of Electrical and Computer Engineering, Georgia Institute of Technology, Atlanta,
% GA, 30332 USA
%\vskip -1\baselineskip plus -1fil

%\begin{IEEEbiographynophoto}{John Dose}
%is with the Department of Electrical and Computer Engineering, Georgia %Institute of Technology, Atlanta, GA, 30332 USA
%\vskip -2.5\baselineskip plus -1fil

%\vfill
\section*{Biographies}
\begin{small}
\textbf{Hang Zou} (hang.zou@tii.ae) is a Researcher at Technology Innovation Institute, UAE. %He holds a Ph.D. in Communications and Networks from Paris-Saclay University. %He is  an IEEE Member.

\textbf{Qiyang Zhao} (qiyang.zhao@tii.ae) is a Lead Researcher at Technology Innovation Institute, UAE. 
%He holds a Ph.D. in Electronic Engineering from University of York. 

\textbf{Lina Bariah} (lina.bariah@tii.ae) is a Senior Researcher at Technology Innovation Institute, UAE. %She holds a Ph.D. in Wireless Communications from Khalifa University. %She is a Senior Member of IEEE. 

\textbf{Mehdi Bennis} (mehdi.bennis@oulu.fi) is a Professor at University of Oulu, Finland.

\textbf{M{\'e}rouane Debbah} (merouane.debbah@ku.ac.ae) is a Professor at Khalifa University, UAE. %He is an IEEE Fellow, a WWRF Fellow, and a Eurasip Fellow. 

% \textbf{Yu Tian} (Yu.Tian@tii.ae) is a Researcher at Technology Innovation Institute, UAE. %He holds a Ph.D. in Electrical and Computer Engineering from  King Abdullah University of Science and Technology.
%\textbf{Faouzi Bader} (carlos-faouzi.bader@tii.ae) is the Director of Telecom Unit at Technology Innovation Institute. %He holds a Ph.D. in Telecom from Universidad Polit{\'e}cnica de Madrid. 
\end{small}

\end{document}